\documentclass[fleqn,10pt]{wlscirep}
\usepackage[utf8]{inputenc}
\usepackage[T1]{fontenc}

\usepackage{upgreek}

\title{Colossal magneto-excitonic effects in 2D van der Waals magnetic semiconductor CrSBr}

\author[1]{R. Komar}
\author[1,*]{M. Goryca}
\author[1,2]{A. Łopion}
\author[3]{M. Rybak}
\author[1,3]{T. Woźniak}
\author[1]{M. Raczyński}
\author[1]{K. M. Gałazka}
\author[4]{K. Mosina}
\author[4]{A. Söll}
\author[4]{Z. Sofer}
\author[1]{W. Pacuski}
\author[5]{C. Faugeras}
\author[1]{M. Birowska}
\author[1,**]{P. Kossacki}
\author[1]{T. Kazimierczuk}
\affil[1]{Faculty of Physics, University of Warsaw, Warsaw, 02-093 Poland}
\affil[2]{Institute of Physics, University of Muenster, Muenster, 48149, Germany}
\affil[3]{Faculty of Fundamental Problems of Technology, Wrocław University of Science and Technology, Wrocław, 50-370 Poland}
\affil[4]{Department of Inorganic Chemistry, University of Chemistry and Technology Prague, Technicka 5, 166 28 Prague 6, Czech Republic}
\affil[5]{LNCMI, UPR 3228, CNRS, EMFL, Université Grenoble Alpes, 38000 Grenoble, France}
\affil[*]{mgoryca@fuw.edu.pl}
\affil[**]{Piotr.Kossacki@fuw.edu.pl}

\begin{abstract}

2D magnetic semiconductors, which intrinsically couple a rich landscape of magnetic orders with tightly bound electron-hole pairs (excitons), present an exciting platform to investigate the interplay between optical and magnetic phenomena at the atomic scale. In such systems, the strength of magneto-optical effects determines how deeply the magnetic properties can be revealed. Here, we report the observation of remarkably strong magneto-excitonic effects in the 2D magnetic semiconductor CrSBr that allow probing its magnetic order with unprecedented sensitivity. By investigating optical transitions above the fundamental exciton energy, we discover a massive spectral shift approaching 100\,meV under applied magnetic fields -- an order of magnitude larger than previously observed magneto-excitonic responses. Our comprehensive magneto-optical experiments accompanied by detailed DFT calculations indicate the possible origin of the transitions exhibiting such intriguing behavior. These findings open avenues for exploiting magneto-excitonic phenomena at newly accessible regimes, enabling novel opto-spintronic applications previously limited by weak magnetic responses.

\end{abstract}
\begin{document}

\flushbottom
\maketitle

\thispagestyle{empty}

\section*{Introduction}

The ability to control the spin degree of freedom of charge carriers is a central pursuit in the rapidly evolving field of spintronics. The coupling between band carriers (electrons or holes) and localized magnetic moments in magnetic semiconductors provides a powerful means to achieve such control, giving rise to intriguing phenomena such as the giant Zeeman effect and spin-polarized carrier transport. Among the plethora of material platforms harnessing this coupling, the recently emerged two-dimensional (2D) magnetic semiconductors \cite{huangLayerdependentFerromagnetismVan2017,gongDiscoveryIntrinsicFerromagnetism2017,cri3_cai, leeIsingTypeMagneticOrdering2016}, wherein magnetic ions are incorporated into an atomically thin semiconductor host, have ushered in a new frontier of opportunities. These 2D systems offer a unique opportunity to explore the interplay between charge, spin, and magnetism at the ultimate scaling limit. One of the most captivating aspects of 2D magnetic semiconductors is the potential for exceptionally strong coupling between the band carriers and the localized d-electrons of the magnetic ions, facilitated by the reduced dimensionality and enhanced exchange interactions. 

The ability to detect and characterize magnetic order in these 2D systems is crucial for understanding the underlying physics and developing spintronic applications. It is, however, experimentally challenging, fueling the development of specialized experimental techniques such as NV-center magnetometry. Optical probes, often preferred due to the possibility of seamless integration with more complex experimental platforms, unfortunately lack universal signatures of magnetic order. Depending on the material, the magnetic properties may manifest as a rotation of the polarization in magneto-optical Kerr effect\cite{Huang2017}, a shift in the Raman-active magnon modes\cite{Pawbake2023}, or a modulation of the polarization degree of the photoluminescence\cite{seylerLigandfieldHelicalLuminescence2018}.

In this respect, the air-stable\cite{air-stable} chromium sulfide bromide (CrSBr) stands out due to particularly clear effects of the magnetic order in the optical spectrum \cite{wilson2021interlayer}. In contrast to, e.g., earlier studied chromium trihalides, the excitons in CrSBr are of Wannier character \cite{wilson2021interlayer} and therefore easier affected by the magnetic order of the sample. The magnetic field dependence is strong enough that it allows for optically observing magnon excitations (spin-waves) in the material in time-resolved experiments\cite{magnons-c,magnons-ab}.

In this work, we report on the observation of unprecedentedly strong magneto-excitonic effects in CrSBr, surpassing the response exhibited by any previously studied 2D semiconductor system. In contrast to prior studies, we focus on the unexplored spectral range above the energy of fundamental excitonic transitions. We show that transitions in this energy range exhibit significantly stronger sensitivity to the magnetic order, manifesting as a colossal spectral shift approaching 100\,meV, as evidenced by our comprehensive magneto-optical experiments with the magnetic field oriented along all three crystallographic directions. Those experimental findings, substantiated by state-of-the-art density functional theory (DFT) calculations, enable us to discuss the possible origins of the observed extraordinary field-dependent behavior. Our results pave the path towards harnessing magneto-excitonic phenomena with an unprecedented level of sensitivity, unlocking a realm of novel experiments and applications that have been constrained by the limited magnetic field responses of conventional systems.

\section*{Basic properties}


As a direct-gap van der Waals semiconductor, CrSBr exhibits prominent excitonic resonances visible both in reflectivity and in photoluminescence spectra. The  orthorhombic crystalline symmetry\cite{magnetic_order} results in strong anisotropy of direct-band gap excitations in the near-infrared energy range\cite{wilson2021interlayer,magnons-c}. Showcasing the intriguing coupling of electronic and magnetic properties, the energy of these excitonic resonances is notably influenced by the magnetic ordering between the layers \cite{ wilson2021interlayer}. Below the critical temperature of 132 K, CrSBr is an A-type antiferromagnet \cite{goser1990magnetic,magnetic_order,layer_orient,boix2022probing}, wherein the spin orientation alternates from layer to layer, with consistent ferromagnetic (FM) intralayer ordering. With that, CrSBr offers three magnetic phase transitions \cite{magnetic_order, air-stable, magnetores}.

\begin{figure}[ht]
\centering
\includegraphics[scale=1]{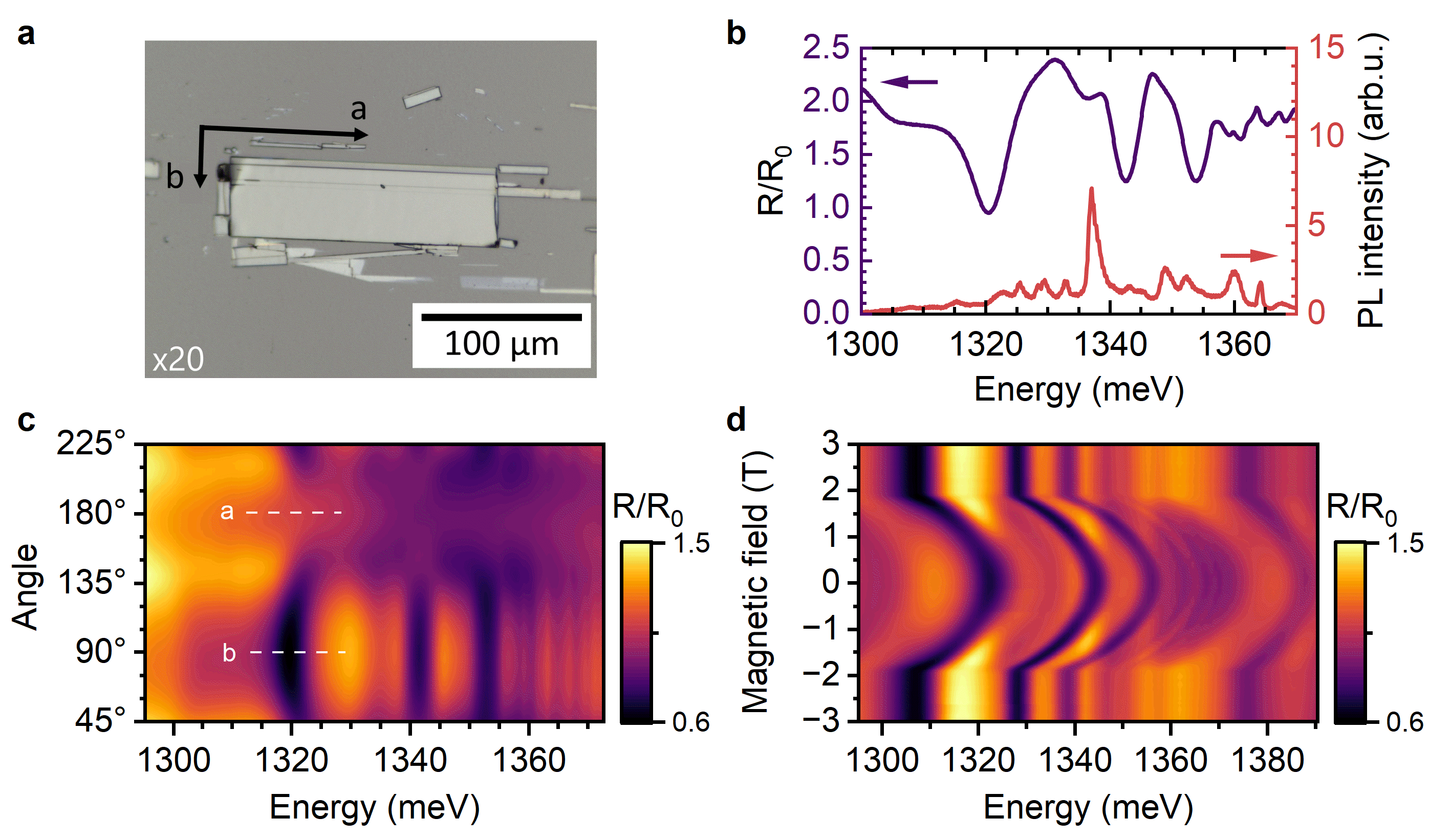}
\caption{\textbf{a}, Optical microscope snapshot of a bulk CrSBr flake with crystallographic in-plane axes marked. The thickness of this flake was checked with the use of atomic force microscopy to be around 700\,nm. \textbf{b}. Reflectance and photoluminescence of a chosen bulk flake. \textbf{c}, False-color plot of reflectance as a function of photon energy (horizontal) and light polarisation angle with respect to the crystal \textit{a} axis (vertical). \textbf{d}, Magnetoreflectance false-color plot of fundamental excitonic states with the magnetic field applied in the Faraday configuration (magnetic field applied along \textit{c}-axis). The reference spectrum $R_0$ is measured on the Si substrate.}
\label{fig:0}
\end{figure}

We begin our investigation by confirming those basic properties of CrSBr and verifying the magneto-excitonic effects reported in previous studies\cite{wilson2021interlayer, dirnberger2023magneto}. The sample was prepared through mechanical exfoliation of bulk CrSBr material onto a Si (111) substrate. The sample contains several rectangular flakes of various thicknesses, in the range of hundreds of nanometers. An exemplary flake is shown in Fig. \ref{fig:0}a. These flakes were characterized using photoluminescence (PL) and reflectance measurements performed at low temperatures (5\,K). The near-infrared optical spectra of each studied flake, both in PL and reflectance, exhibit several spectral features that can be identified as exciton-polaritons with energies ranging from 1.3 to 1.4 eV (see Fig. \ref{fig:0}{b}), which is typical for bulk-like CrSBr flakes \cite{ultra-strong_coup}. As demonstrated in Fig. \ref{fig:0}c, all the observed excitonic features are linearly polarized along the \textit{easy} \textit{b}-axis of the flake, which is a consequence of the highly anisotropic crystal structure of CrSBr \cite{wilson2021interlayer}.

By subjecting the flakes to an external magnetic field along the \textit{c}-axis, we corroborate the strong magneto-excitonic effects, as presented in Fig. \ref{fig:0}{d}. Upon increasing the magnetic field up to the complete interlayer FM phase transition near $B=2$\,T, the excitonic features exhibit a significant redshift of approximately 20\,meV. Above the point of magnetization saturation, the field dependence becomes much weaker, unambiguously demonstrating that the observed changes originate from the inter-layer antiferromagnetic ordering rather than the direct interaction of the excitons with an external magnetic field. A detailed discussion of the underlying mechanism can be found in Ref. \citeonline{wilson2021interlayer} and is also recapitulated in the following section.

\section*{Ultra strong magneto-electronic coupling in higher-bands excitonic states}

\begin{figure}[ht]
\centering
\includegraphics[scale=1]{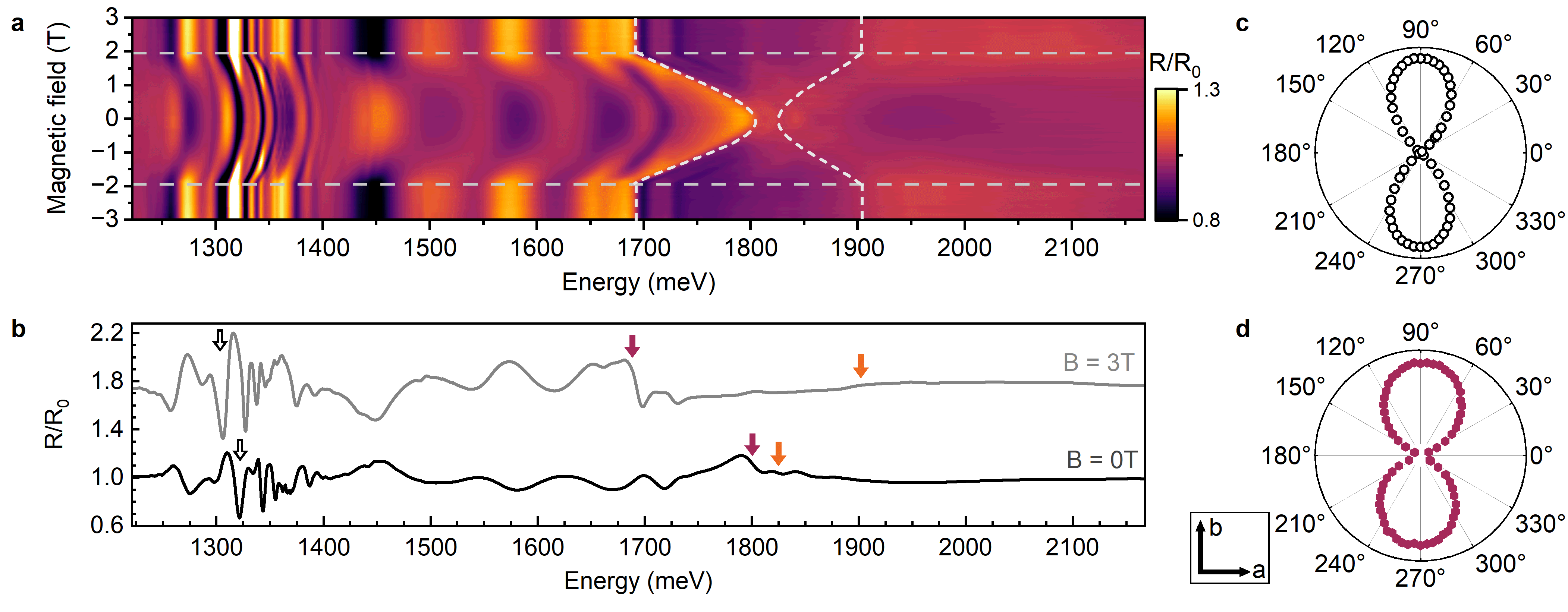}
\caption{\textbf{a}, False-color plot depicting the magnetoreflectance as a function of photon energy (horizontal axis) and external magnetic field (vertical axis) applied along the\textit{c-axis} of the flake. The horizontal dashed lines highlight the magnetic field values at which magnetization saturates due to complete alignment of the spins with the external magnetic field. The white dashed lines serve as guides to the eye for the $\Omega_3$ and $\Omega_4$ excitons discussed in the text. An overlay-free version of this plot can be found in the Supplementary Material. \textbf{b}, Comparison of reflectance spectra acquired in the absence of an external magnetic field and with a 3\,T field applied along the \textit{c-axis}. The arrows indicate fitted positions of: $\Omega_{1/2}$ (dark outline) and $\Omega_3$ (crimson) and approximated positions of $\Omega_4$ (orange).  \textbf{c-d} Anisotropy of the fitted absorption state intensities as a function of light polarization angle relative to the \textit{a-axis} for: \textbf{c}, fundamental excitonic state $\Omega_{1/2}$ ($1.3$\,eV), with data mirrored across a horizontal line; \textbf{d}, reported states marked with arrows on the spectra in panel \textbf{b}. All measurements (\textbf{a}–\textbf{d}) were performed at a sample temperature of $T=5$\,K.}
\label{fig:1}
\end{figure}

While previous magneto-optical studies focused on the fundamental excitons in the range between 1.3 and 1.4\,eV, numerical calculations \cite{theory_HE,1D} predict the existence of higher-energy bands with comparable oscillator strengths. To investigate such transitions, we performed a series of optical experiments over a wide energy range from 1.2\,eV to 2.2\,eV. The reflectance spectrum within this range reveals the presence of additional excitonic resonances in the energy range of 1.7 to 1.9\,eV, as can be observed in Fig. \ref{fig:1}a and is indicated by color arrows in Fig. \ref{fig:1}b. To unambiguously confirm the nature of these transitions, we analyzed their polarization dependence. As depicted in the polar plots in Figs. \ref{fig:1}c and \ref{fig:1}d, these high-energy spectral features exhibit the same linear polarization as the fundamental excitons, strictly aligned with the b-axis of the crystal.

The most striking property of these excitonic transitions is its enormous dependency on the external magnetic field, as uncovered by our magnetoreflectance measurements (Fig. \ref{fig:1}{a}).
At low temperatures, the reported states offer energy tunability of more than 80\,meV over the span of 2\,T for the magnetic field along the \textit{c}-axis. Apart from the vastly larger energy scale, the overall shape of the shift in the magnetic field is similar to the one observed for the fundamental excitons in 1.3-1.4\,eV, thus suggesting a similar underlying mechanism.

The core concept of the explanation of those observations is that the exciton spins are locally aligned to the magnetization of the particular monolayer\cite{wilson2021interlayer}. At $B=0$, the neighboring layers are oriented along the easy axis (\textit{b}-axis), anti-parallel between the nearest-neighbor layers. Due to this anti-parallel orientation, the hybridization of the excitons between the layers does not occur. As we increase the magnetic field along the \textit{c}-axis, the magnetization of each monolayer becomes canted toward the direction of the field. The $c$-projection of the neighboring layer magnetization follows $M \propto B_\mathrm{ext}$ , and thus the degree of inter-layer hybridization (and correspondingly -- lowering of the exciton energy) follows $\Delta E\propto B_\mathrm{ext}^2$. Around $B=2$\,T value the magnetization of each layer becomes fully oriented along the external magnetic field, and thus becomes saturated at its maximum value. Above that point, the relative magnetization between the neighboring layers does not change anymore, and the only remaining factor affecting the exciton energy is a regular excitonic Zeeman effect, much weaker than the inter-layer hybridization effect described above. 

Such a mechanism corresponds well with the behavior illustrated in Fig. \ref{fig:1}{a}: below the transition point of 2\,T the exciton energy dependence on the magnetic field is quadratic, and above 2\,T the exciton energy stays almost constant. As we show in the further section, the energy dependence on the magnetic field below 2\,T is so strong that it actually deviates from strictly quadratic dependence.

\begin{figure}[ht]
\centering
\includegraphics[scale=1]{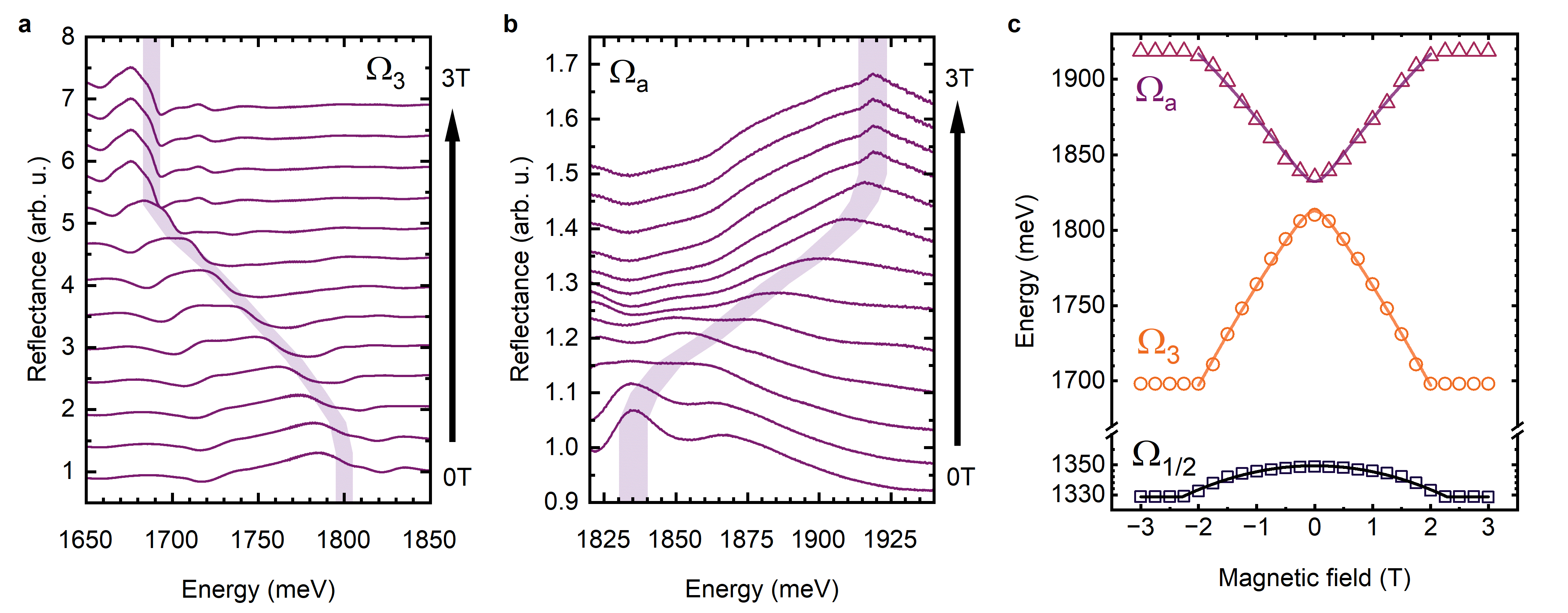}
\caption{\textbf{a-b} Magnetic field dependence of reflectance spectra in energy range corresponding to \textbf{a}, $\Omega_3$, and  \textbf{b}, $\Omega_4$, transitions. Shaded areas serve as a guide to the eye indicating the position of the transition. \textbf{c}, Dependence of extracted transition energy for all analyzed excitons versus magnetic field along the \textit{c}-axis. The measurement temperature was T = 5\,K.}
\label{fig:4th}
\end{figure}

Moreover, careful analysis of the gathered data (see Fig. \ref{fig:4th}{a,b}) reveals the existence of yet another excitonic transition with qualitatively different behavior. Specifically, at $B=0$\,T the transition occurs at about 1.84\,eV and with increasing magnetic field is shifted towards \emph{higher} energy. Again, the shift is almost quadratic as a function of external magnetic field up to $B\approx 2$\,T, reaching $\Delta E\approx 85$\,meV.
To enhance the visibility of the discussed excitonic transition we prepared a sample with flakes transferred on the CdTe-based heterostructure to provide better interference conditions. The magnetoreflectance map of that sample is included in the Supplementary Material.

The exact identity of the observed higher-energy excitons remains unknown. Theoretical studies by Qian \textit{et al.} \cite{theory_HE} as well as by Klein \textit{et al.} \cite{1D} predict two states originating from higher bands in a similar energy range, but with no magnetic field in the simulations the attribution of the observed peaks is only tentative. Nevertheless, in the following discussion, we will adopt the notation  from Ref. \citeonline{1D}, denoting the fundamental excitons (in the 1.3-1.4\,eV range) as $\Omega_{1/2}$, the exciton red-shifted by magnetic field from 1.8\,eV to 1.7\,eV as $\Omega_3$, and the exciton blue-shifted by magnetic field from 1.82\,eV to 1.9\,eV as $\Omega_\textsc{4}$.

Mirror-like magnetooptical behavior of $\Omega_\textsc{3}$ and $\Omega_\textsc{4}$ excitons in Fig. \ref{fig:1}{a} suggests a~simple phenomenological description of these states in terms of field-induced coupling. In such a picture, by diagonalizing a 2x2 Hamiltonian we derive the following expressions for the field dependence in the regime below the saturation point:
\begin{equation}
\begin{array}{cc}
E_3(B) \, = \, \frac{1}{2} \left( E_3 + E_4  - \sqrt{\left(E_4-E_3\right)^2 + \left(\alpha B\right)^2 }\right) - \beta B^2  , &
E_4(B) \, = \, \frac{1}{2} \left( E_3 + E_4  + \sqrt{\left(E_4-E_3\right)^2 + \left(\alpha B\right)^2 }\right) - \beta B^2  , \\
\end{array}
\end{equation}
where $E_3$ and $E_4$ is the zero-field energy of $\Omega_3$ and $\Omega_4$ exciton, respectively, $\alpha B$  is the field-induced coupling between $\Omega_3$ and $\Omega_4$ excitons, and term of $-\beta B^2$ originates from coupling with remote bands, similarly as in the field dependence of $\Omega_{1/2}$ excitons. Values calculated using these formulas are included in Fig. \ref{fig:4th}{c} as thin lines. Notably, they fit significantly better the experimental energy shifts than simple quadratic dependence. Combined with the fact of natural emergence of the \emph{inverted} energy shift of $\Omega_4$, the proposed phenomenological description seems to capture the main underlying mechanism of the observed giant sensitivity of the $\Omega_3$ and $\Omega_4$ excitons on magnetic field.

In order to further test the link between the layer magnetization and the behavior of $\Omega_3$ and $\Omega_4$ excitons we have performed additional measurements with external magnetic field applied along the \textit{a}- or \textit{b}-axis. In both cases the qualitative behavior of the $\Omega_3$ and $\Omega_4$ excitons was consistent with the changes reported earlier for fundamental $\Omega_{1/2}$ excitons in such field geometry \cite{wilson2021interlayer}. Applying the magnetic field along the easy-axis (\textit{b}-axis) leads to sudden jump of the exciton energies at field of about $B=0.35$\,T. The full magneto-reflectance false-color plots for these directions are included in the Supplementary Material.

\section*{DFT Calculation of the Band Structure}

\begin{figure}
  \vspace{0pt}
  \centering
 \includegraphics[scale=0.58]{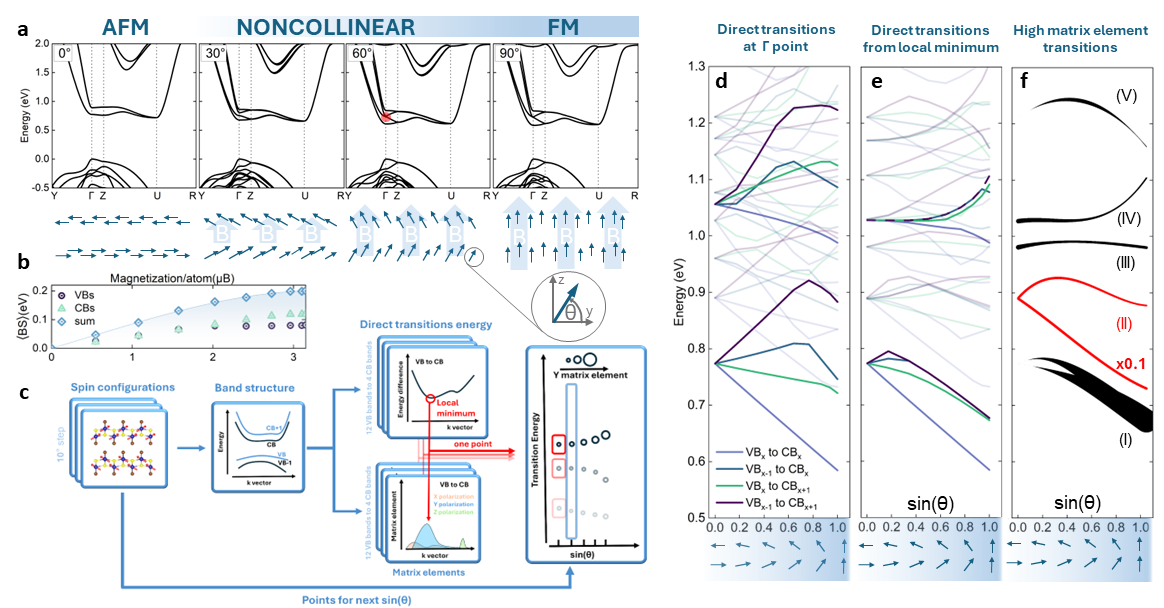}
\vspace{0pt}
\caption{a) Electronic bandstructures obtained within DFT+U+SOC approach for particular $\theta$ angles of bulk CrSBr, with a schematic arrangements of the spins from adjacent layers. The zoom out schematically defines  the $\theta$ angle of spin rotations, under the impact of the out-of plane external magnetic field.  b) The average  spin band splitting (BS)  of the 12 VBs and 4 CBs  at $\Gamma$ point, without inclusion of spin degeneracy in A-AFM case (0 magnetization), presented separately for VBs, CBs and their sum. c) Flowchart of computation scheme to obtain the electronic transitions in respect to the strength of the magnetic field (see detailed description in main text). d) The evolution of the direct transitions in respect to the strength of magnetic field ($\sim sin(\theta)$ in simulations) for transitions located at: d) $\Gamma$ point,  e) local minimum along $\Gamma-Z$ direction, with particular low lying transitions (centered $\sim0.8$\,eV)  and higher electronic transitions (centered $\sim1.1$\,eV) denoted in bold,  and f) for transitions exhibiting the highest dipole matrix elements, where the thickness of each point indicates the strength of the \textit{y} component of linearly polarized light. }
\label{fig:theory}
\end{figure}

Contributing factors to the remarkably strong dependence of the $\Omega_3$ and $\Omega_4$ excitons' energy on the external magnetic field could, in principle, include also field-induced modifications to the electronic structure of the host material. Thus, to provide more insight into the possible physical origins underlying the observed colossal energy shifts of the $\Omega_\textsc{3}$ and $\Omega_\textsc{4}$ transitions, we have undertaken state-of-the-art Density Functional Theory (DFT) calculations to investigate the bulk CrSBr band structure. We systematically examine the evolution of the interband transitions upon rotation of Cr spins (measured by $\theta$ angle defined in Fig. \ref{fig:theory}(a)). We consider the noncollinear rotations of the spins in adjacent layers that mimic the effect of out-of-plane external magnetic field ($\vec{B}_{\perp}$) in experiments so that $\theta$ reflects the strength of the field between two extreme cases: A-AFM ($\theta=0^{\circ}$) and FM ($\theta=90^{\circ}$) (which correspond to the 0T and 2T in the experiment, respectively).
 
We start by performing high-throughput calculations considering 12 valence bands (VB) and  4 conduction bands (CB),  that yield 48 plausible direct electronic transitions for each spin orientation and each k-point. We consider 10 different spin orientations (defined by $\theta$) and various k-points selected at high symmetry lines, in contrast to previous reports \cite{} where transitions only at the $\Gamma$ points have been considered. The choice of k-paths is guided by earlier theoretical considerations outlined in ref. \cite{D3TC01216F}. We adopt the calculation strategy described in detail in Supplementary Material and presented schematically in the flowchart (Fig. \ref{fig:theory}(c)), confining our theoretical approach to first principles calculations within the PBE+U+SOC framework. Consequently, we focus on the energy shifts of the transitions upon the rotation of the Cr spins, rather than the absolute values, neglecting the exciton binding energy of particular transitions. For all considered direct interband transitions we calculate the dipole strength and focus on those transitions which couple to the \textit{y} component of the linearly polarized light, aligning with the experimental observations. As a result, we construct a plot shown in Fig. \ref{fig:theory}f that presents the energy shifts of the interband transitions as a function of the Cr spin angle $\theta$. Note that the plot includes first few electronic transitions exhibiting the largest dipole strength, hence the most significant optical activity. Comprehensive details can be found in the Supplementary Material.


Our results unveil strong dependence of the electronic structure on the rotation angle $\theta$ (see Fig. \ref{fig:theory}(a) and SM). Notably, giant spin directional band splitting dependence have been reported previously in monolayer of CrI$_3$, whereas in bulk form is quenched \cite{Jiang2018}. In the case of  FM state the spin  degeneracy of the bands is lifted compared to A-AFM case, in line with previous reports \cite{}. In fact, the band spin splitting occurs for \textit{any} rotation exhibiting net magnetization and depends strongly on k-point selection (see Fig \ref{fig:theory}(b)). In particular, two lowest lying CBs in the A-AFM state are split into 4 CBs for $\theta\neq0$, resulting in the increased number of transitions. Notably, a band splitting is observed with emergence of FM state also along in-plane directions as we previously reported in  \cite{D3TC01216F} (see SM therein), irrespective to the direction of the net magnetization. 

Although our single-particle approach cannot \textit{fully} account for the general field-induced behavior of the $\Omega_{1/2}$, $\Omega_3$, and $\Omega_4$ transitions, it correctly captures many observed features Some of the strongest higher-lying transitions in our simulations exhibit colossal red- or blueshifts approaching or even exceeding 100 meV (e.g., the lower of the two (II) transitions or the (IV) transition in Fig. \ref{fig:theory}(f)). In particular, the pair of the (II) transitions, with the upper one initially blueshifting, exhibit behavior qualitatively consistent with previously discussed simple phenomenological description in terms of field-induced coupling, given the existence of additional source of the zero-field splitting not included in our DFT calculations. However, these calculated colossal shifts are accompanied by comparable shifts obtained for the lowest-lying transitions, whereas the experimental results demonstrate an order of magnitude smaller shift for the $\Omega_{1/2}$ transition. Additionally, the energy separation between the calculated lowest-lying transitions and those that could potentially correspond to $\Omega_3$ and $\Omega_4$ does not match the experimental observations. This discrepancy clearly indicates that the large exciton binding energy, not accounted for in our single-particle approach, may play a crucial role in understanding the observed landscape of field-dependent transitions.

The binding energy for the fundamental transition has been reported to be 870 meV, whereas for higher transitions, it has been calculated to be approximately 500 meV for a CrSBr monolayer in the FM state \cite{PhysRevResearch.5.033143}. This fact alone could explain the mismatch between the observed and calculated energy separations between the selected states. Moreover, the exciton binding energy for various spin configurations can differ, as reported for other intrinsic magnetic layered materials \cite{PhysRevB.109.054426,PhysRevB.103.L121108}. Indeed, our calculations reveal that the curvature of the electronic bands strongly depends on the magnetic field strength (see Supplementary Material), corroborating the picture of field-dependent exciton binding energies. This effect may have the most profound impact on the observed $\Omega_{1/2}$ shift, as the binding energies of the lowest-lying states are the largest, partially compensating the field-induced changes in the electronic band structure. However, to elucidate the exact role of this effect, further many-body calculations are required, which remains beyond the scope of this work.

\section*{Effect of the temperature}

\begin{figure}[ht]
\centering
\includegraphics{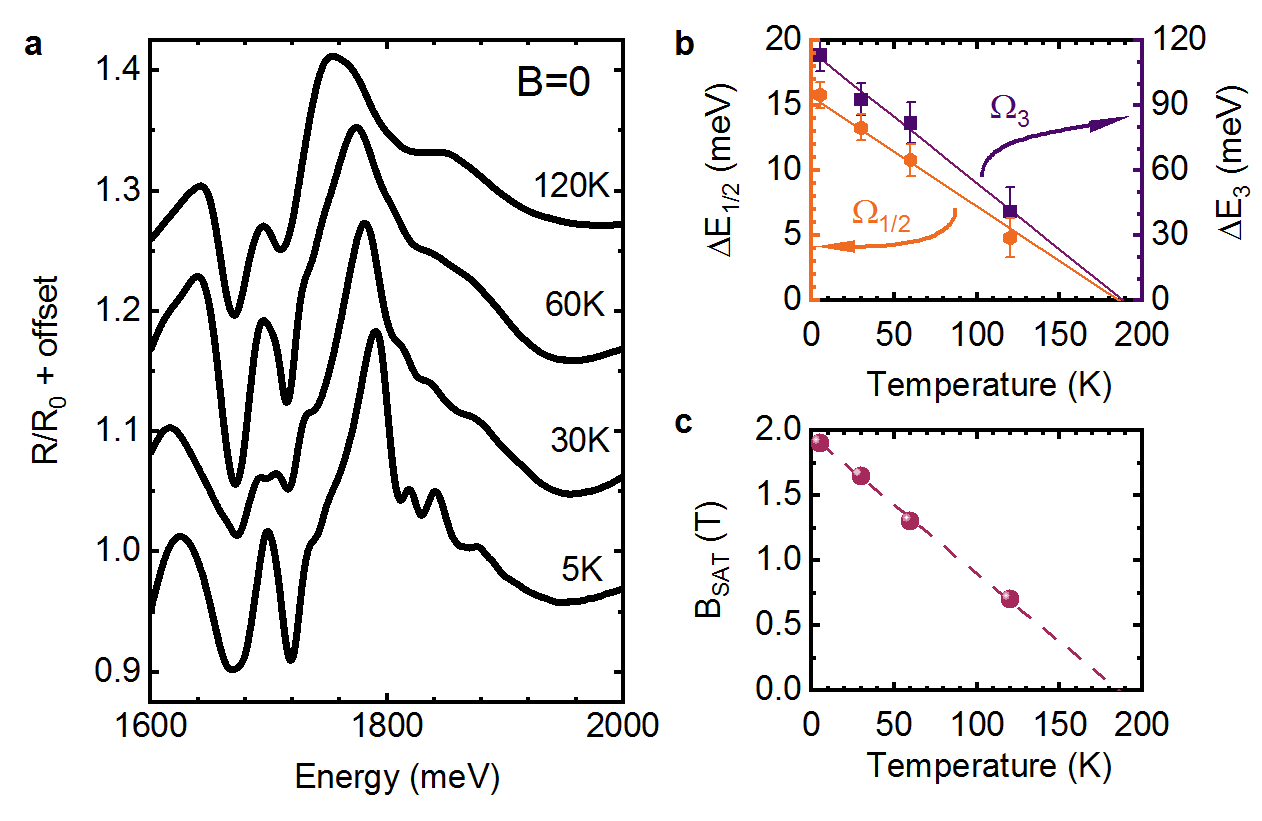}
\caption{
\textbf{a}, Comparison of reflectance spectra near $\Omega_3$ exciton for different temperatures.
\textbf{b}, Total energy shifts ($\Delta E$) in magnetic field along \textit{c-axis} of $\Omega_{1/2}$ and $\Omega_3$ excitons versus temperature together with the Linear extrapolation.
\textbf{c},  Magnetic field corresponding to the magnetization saturation, common for all observed excitonic states. The linear extrapolation gives critical temperature of about 185\,K.
\label{fig:2}}
\end{figure}

As a final test of the origin of the intriguing behavior of the $\Omega_3$ and $\Omega_4$ excitonic resonances and its connection with the magnetic order within the crystal lattice, we harness yet another degree of freedom at our disposal: temperature. While all our results discussed so far were obtained at cryogenic conditions ($T=5$\,K), the antiferromagnetic phase of CrSBr is known to persist, albeit in a more fragile state, up to the Néel temperature of $T=132$\,K \cite{goser1990magnetic}. Seizing this opportunity, we perform additional magneto-reflectance measurements across an elevated temperature range. The results of this investigation is shown in Figs. \ref{fig:2}{a}-{c}.

Upon increasing temperature, we observed gradual changes in the energy scale as well as the value of the magnetic field of the saturation point. Within experimental uncertainty, both these parameters decrease linearly with the increasing temperature, which is shown in Figs. \ref{fig:2}{b} and \ref{fig:2}{c}, respectively. Such a behavior is distinctively different than, e.g., the strength of the second-harmonic generation signal in Ref. \cite{magnetic_order}, which was shown to follow $(1-T/T_c)^\beta$ with critical exponent $\beta=0.36$. On top of that, the critical temperature obtained by extrapolating the data in Figs. \ref{fig:2}{b} and \ref{fig:2}{c} is about 185\,K, which is significantly above the Néel temperature (132\,K in bulk). The temperature of 185\,K extrapolated from our experiments is much closer to the third critical temperature found in Refs. \citeonline{liu2022three,dirnberger2023magneto}, which corresponds to the intermediate ferromagnetic phase, i.e., when the inter-layer antiferromagnetic coupling is already overcame by the temperature, but the intra-layer ferromagnetic order is present separately in each layer. This observation is reinforced by an argument, that the magnetization canting behavior is determined by the interplay between magnetocrystalline anisotropy and the external magnetic field, both of which are independent on inter-layer ordering, which in turn is pertinent for A-type antiferromagnetism. The disappearance of the inter-layer ordering upon approaching the Néel temperature is however a factor contributing to the gradual disappearing of the excitonic lines at elevated temperatures. As shown in Fig. \ref{fig:2}{a}, the excitonic features in the optical spectra become significantly broader at temperatures above ~100\,K, which is to be expected due to increasing randomness of relative inter-layer orientations.

\section*{Summary}


In summary, we have observed unprecedentedly strong magneto-excitonic effects in the CrSBr that surpass the response of any previously studied 2D semiconductor system. By exploring the unexplored spectral range above the energy of fundamental excitonic transitions, we have demonstrated that transitions in this higher energy range exhibit a significantly stronger sensitivity to the magnetic order. These higher energy transitions show identical anisotropy and similar gradual field dependence as the fundamental excitonic transitions, but offer an order of magnitude larger field-induced spectral shifts, approaching 100\,meV. Our experimental findings are corroborated by state-of-the-art DFT calculations that indicate that a possible origin of the colossal spectral shifts are the field-induced changes in the electronic band structure, which in the case of the lowest energy transitions might be partially compensated by changes in the exciton binding energy. The discovery of such colossal magneto-excitonic effects in CrSBr represents a significant advancement in the understanding of magneto-optical coupling in 2D magnetic semiconductors and its exploration with unprecedented sensitivity levels, overcoming limitations imposed by the relatively weak magnetic responses in conventional systems.

\section*{Acknowledgements}
This work was supported by National Science Centre, Poland under projects 2020/38/E/ST3/00364 and 2020/39/B/ST3/03251. M.G. received support from the Polish National Agency for Academic Exchange within Polish Returns program under Grant No. PPN/PPO/2020/1/00030. Z.S. was supported by ERC-CZ program (project LL2101) from Ministry of Education Youth and Sports (MEYS) and used large infrastructure from project  reg. No. CZ.02.1.01/0.0/0.0/15\_003/0000444 financed by the EFRR. A. S. was financially supported by the Marie Curie Sklodowska ITN
 network “2-Exciting” (grant no. 956813).
 
 \vspace{3mm}
 
 \textbf{Author contributions:} R.K., K.M, A.S. Z.S, W.P., and C.F. prepared the materials and fabricated the samples. R.K., A.Ł, and T.K. carried out the experiments. R.K., A.Ł., M.G., P.K., and T.K. analyzed the experimental data. M.R., T.W., and M.B. performed the DFT calculations. R.K., M.G., P.K. and T.K. wrote the paper. All authors contributed to the final manuscript.

\bibliography{literatura}

\end{document}


\flushbottom

\begin{abstract}
\end{abstract}
\maketitle

\tableofcontents
\clearpage

\section{Computational Details}

\subsection{Excitonic calculations (TDDFT+mBSE)}

To accurately reproduce the optical spectra and capture excitonic effects arising from electron--hole interactions, we employed a two-step computational strategy combining ground-state hybrid-functional calculations with excited-state TDDFT+mBSE analysis, as implemented in \textsc{VASP} (version 6.4.3)  \cite{KRESSE199615}.

\textbf{Ground-State Calculations}. The ground-state electronic structure was obtained within the projector augmented-wave (PAW) formalism \cite{PhysRevB.59.1758} using the Perdew--Burke--Ernzerhof (PBE) exchange--correlation functional \cite{PhysRevB.54.11169,KRESSE199615}. Experimental lattice parameters were adopted (\(a = 3.5761~\text{\AA}\), \(b = 4.7859~\text{\AA}\), \(c = 7.9838~\text{\AA}\))~\cite{Lopez-Paz2022}. To describe the A-type antiferromagnetic (A-AFM) stacking of layers, the out-of-plane lattice vector was doubled, ensuring alternating spin alignment between adjacent layers. Atomic coordinates were relaxed until residual Hellmann--Feynman forces were below \(10^{-3}\)~eV~\AA\(^{-1}\), while keeping the lattice constants fixed. The plane-wave kinetic energy cutoff was set to 450~eV, and the Brillouin zone was sampled with a \(16 \times 12 \times 4\) Monkhorst--Pack grid. Electronic occupations were treated with a Gaussian smearing of 0.05~eV, and the total energy convergence threshold was \(10^{-7}\)~eV. These parameters ensured fully converged total energies and forces, and provided sufficiently accurate Kohn–Sham wavefunctions for subsequent evaluation of optical transition matrix elements.
\begin{figure}[!h]
  \vspace{0pt}
  \centering
 \includegraphics[scale=1.2]{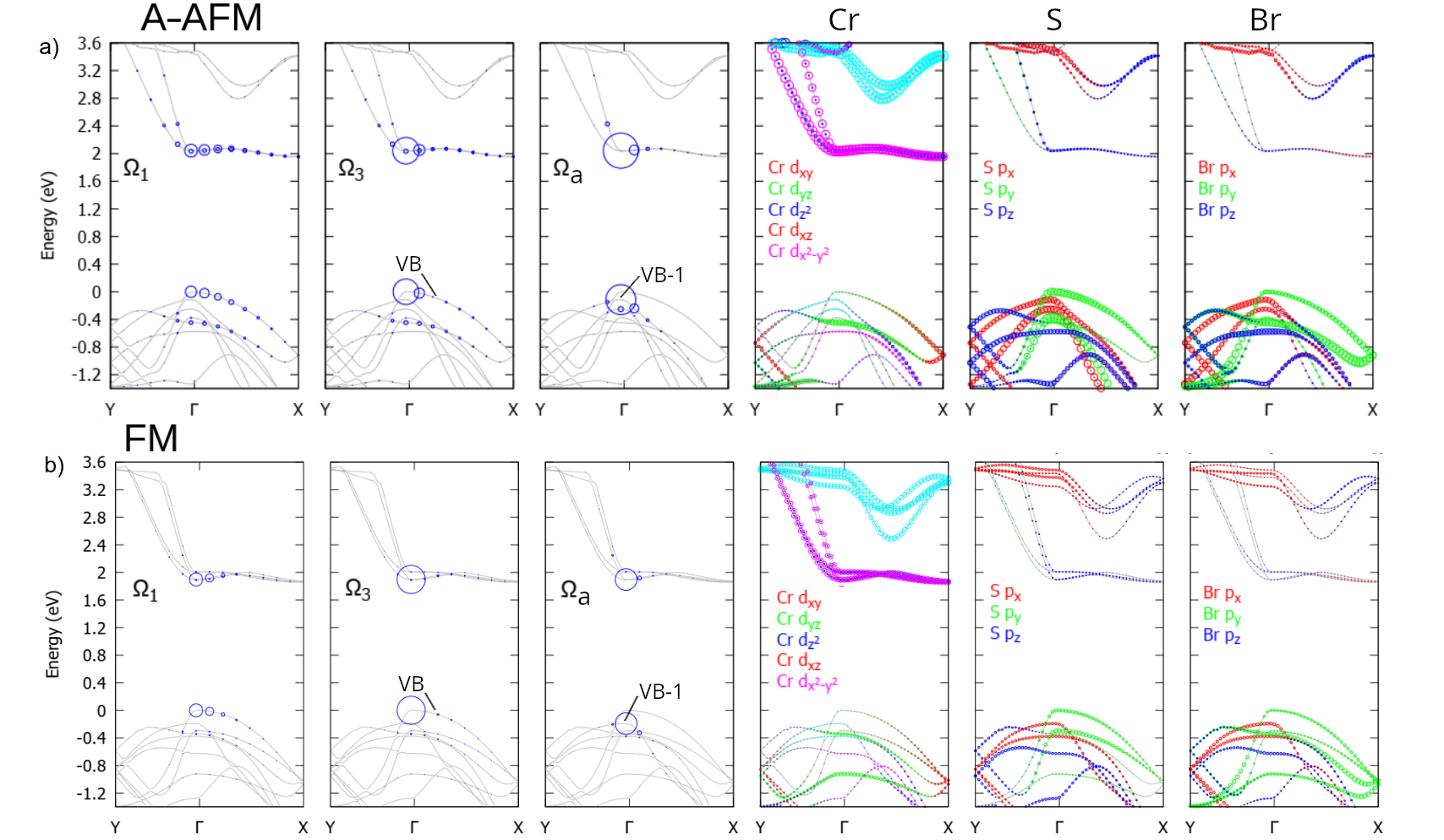}
\vspace{0pt}
\caption{
Band structures obtained using the double-screened hybrid (DSH) functional and excitonic properties calculated within the time-dependent density functional theory combined with the model Bethe--Salpeter equation (TDDFT+mBSE) framework. 
(a,b) The first three panels from the left display the excitonic wavefunctions corresponding to the first bright excitonic states (from left to right: $\Omega_{1/2}$, $\Omega_{3}$, and $\Omega_{a}$), projected onto the band states. 
The size of each circle represents the magnitude of the corresponding contribution to the excitonic wavefunction. 
The last three panels show the orbital-resolved band projections for Cr, S, and Br atomic states. 
Panels (a) and (b) correspond to the A-type antiferromagnetic (A-AFM) and ferromagnetic (FM) phases, respectively, both calculated within the spin-polarized framework.
}
\label{fig:S1}
\end{figure}

\textbf{Double-Screened Hybrid (DSH) Functional}. To obtain an improved description of the electronic band structure and to reproduce quasiparticle-like band gaps, the double-screened hybrid (DSH) functional~\cite{acs.jpclett.8b00919} was employed. The DSH approach extends conventional hybrid functionals by incorporating a physically motivated, dielectric-dependent screening of the Fock exchange interaction. In this framework, the Coulomb potential is separated into short- and long-range components, where the range-separation parameter \(\mu\) defines the spatial extent of the screened exchange. Each part of the exchange interaction is then weighted by an effective screening determined from the material’s electronic dielectric response. The short-range and long-range contributions are screened differently, reflecting local and macroscopic polarization effects, respectively. Two parameters govern the DSH functional \cite{acs.jpclett.8b00919}: (i) the fraction of long-range exact exchange, \(\beta\), and (ii) the range-separation parameter, \(\mu\). In the self-consistent DSH procedure implemented in \textsc{VASP}, these parameters are derived from the system’s own electronic screening without any empirical fitting. Specifically, the macroscopic dielectric constant \(\epsilon_\infty\) is first evaluated within a model random-phase approximation, and the long-range exchange fraction is set to \(\beta = 1/\epsilon_\infty\). The parameter \(\mu\) is subsequently obtained by fitting a model dielectric function, which describes the momentum-dependent screening of the Coulomb interaction. This procedure yields a fully self-consistent dielectric-dependent hybrid functional, where the amount and range of exact exchange are directly linked to the material’s intrinsic screening properties.  In practice, the self-consistent determination of \(\beta\) and \(\mu\) was performed separately for the antiferromagnetic (AFM) and ferromagnetic (FM) configurations, resulting in
\(\beta_{\mathrm{AFM/FM}} = 0.1545/0.1561\) and
\(\mu_{\mathrm{AFM/FM}} = 2.8040/2.7446~\text{\AA}^{-1}\).
These values correspond to physically reasonable dielectric constants consistent with moderately screened exchange. To ensure quantitative agreement with experiment, the final DSH parameters were slightly adjusted so that the resulting DSH band gaps reproduced the experimental quasiparticle band gaps extracted from angle-resolved photoemission spectroscopy (ARPES) \cite{PhysRevB.107.235107}. The resulting DSH-corrected eigenvalues and wavefunctions were subsequently used as input for the optical and excitonic calculations described below.

\textbf{Excitonic and Optical Calculations}
The optical response was computed within time-dependent density functional theory (TDDFT) using the model Bethe--Salpeter equation (mBSE) formalism in the Tamm--Dancoff approximation (TDA), as implemented in \textsc{VASP}. This approach includes both the screened electron--hole attraction and the exchange-type local-field effects, thus accounting for excitonic contributions to the dielectric function. The mBSE Hamiltonian was constructed using 20 occupied and 4 unoccupied bands, while a total of 120 bands were included in the dielectric-function convergence tests. A complex shift of 0.001~eV was applied to simulate Lorentzian broadening of the spectral features in \(\mathrm{Im}[\epsilon(\omega)]\). Because full excitonic calculations for arbitrary magnetic canting angles (\(\theta\)) are computationally demanding, the TDDFT+mBSE analysis was restricted to the AFM (\(B = 0~\mathrm{T}\)) and FM (\(B = 2~\mathrm{T}\)) configurations, each treated in a fully spin-polarized framework to capture magnetically induced optical anisotropy.

\begin{figure}[!h]
  \vspace{0pt}
  \centering
 \includegraphics[scale=1.2]{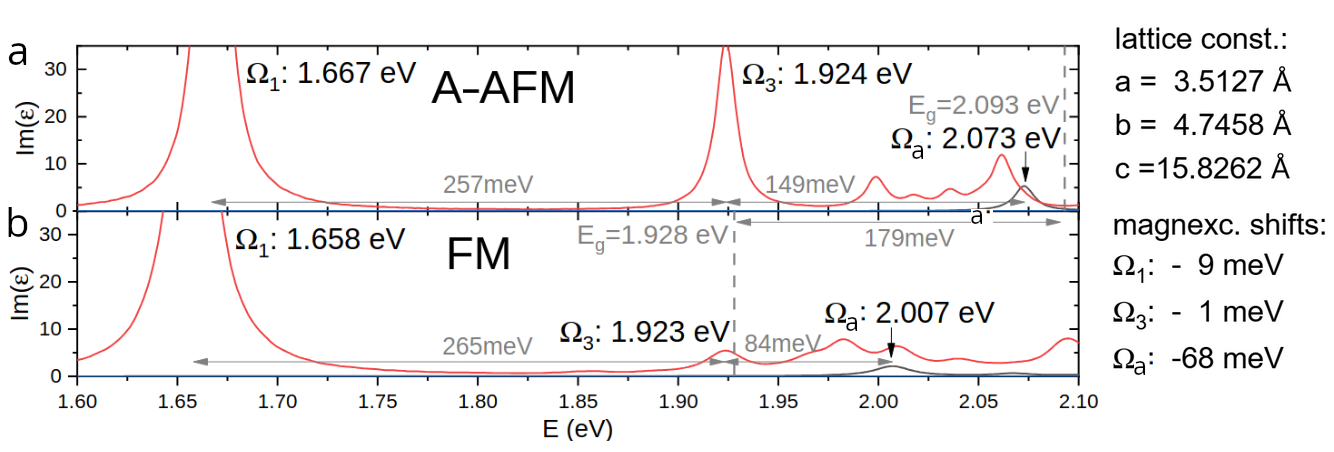}
\vspace{0pt}
\caption{
Imaginary part of the dielectric function for two magnetic orderings of bulk CrSBr: (a) A-type antiferromagnetic (A-AFM) and (b) ferromagnetic (FM) phases.
The spectra were obtained within time-dependent density functional theory combined with the model Bethe--Salpeter equation (TDDFT+mBSE), based on the experimental lattice parameters.
The right panel shows the energy shifts associated with the transition from the A-AFM to the FM phase.
Red and yellow colors indicate the light polarization coupled to the $y$ and $x$ directions, respectively.
}
\label{fig:S2}
\end{figure}

The electronic band structure obtained using the double-screened hybrid (DSH) functional is shown in Fig.\ref{fig:S1}, together with the excitonic wavefunctions projected onto the band states to visualize the dominant valence–conduction transitions contributing to the bright ex citations: $\Omega_{1/2}$, $\Omega_{3}$, and $\Omega_{4}$. The optical absorption spectra, calculated within the TDDFT+mBSE framework for thw magnetic phases A-AFM and FM, are shown in Fig.\ref{fig:S2}. The absorption intensity is proportional to the imaginary part of the macroscopic dielectric function, highlighting the excitonic features and their dependence on magnetic ordering.

Importantly, all considered excitons red-shift when going from A-AFM to FM, but the calculated shifts are more than two times smaller than in experiment. Given that these shifts are only a few tens of meV, they lie within the sensitivity of our workflow to technical parameters such as the lattice constants, k-mesh density, dielectric screening ($\beta$, $\mu$), and the size of the BSE subspace. Consequently, within our present settings, the precise sign and magnitude of the shift cannot be regarded as conclusive.

\subsection{DFT+U calculations}

A fully rigorous treatment of excitonic behavior in the presence of an external magnetic field would require solving the Bethe–Salpeter equation (BSE) with explicit inclusion of noncollinear magnetization and spin–orbit coupling (SOC). Such calculations are computationally demanding for CrSBr due to its large unit cell and the necessity of dense Brillouin-zone sampling. Instead, we adopt a simplified yet physically justified approach: density functional theory plus Hubbard~$U$ (DFT+$U$) calculations including SOC were performed for both the A-type antiferromagnetic (A-AFM, $B=0$~T) and ferromagnetic (FM, $B\!\approx\!2$~T) configurations. The underlying reason is that the dominant magnetic-field dependence of the excitonic spectrum arises from modifications to the electronic band structure—specifically, shifts in band energies and corresponding changes in orbital character. Notably, the excitonic resonances $\Omega_3$ and $\Omega_a$ exhibit mutually orthogonal polarizations, originating from the distinct orbital compositions of the electronic bands that contribute to their respective excitonic wavefunctions. This relationship enables the identification and tracking of the electronic bands associated with specific polarization channels.

The calculations have been conducted within  density functional theory (DFT) using the generalized gradient approximation  as implemented in VASP software \cite{KRESSE199615}. The ion–electron interactions have been described by the projector augmented wave (PAW) method  \cite{Holzwarth2001}. Plane-wave basis cutoff and $\Gamma$ centered Monkhorst-Pack \cite{Monkhorst1976} k-point grid have been set to 500 eV and $10\times6\times1$, respectively. A Gaussian smearing of 0.05 eV was employed for the Brillouin zone (BZ) integration. The interlayer vdW forces have been treated within Grimme scheme \cite{Grimme} using D3 correction \cite{DFT-D3}. The results were obtained  using PBE+U method based on Dudarev's approach \cite{PhysRevB.57.1505}, with the effective on-site Hubbard U parameter ($U_\mathrm{eff} =U-J$, where $J$ has been fixed to $J=1$~eV) assumed for $3d$ orbitals. Note that $U_\mathrm{eff}$ is hereafter denoted as $U$. The impact of U parameter on electronic transitions  have been previously considered in paper \cite{D3TC01216F}.  The SOC within the non-collinear treatment of magnetism  has been taken into account on the top of the PBE+U scheme. The position of the atoms and unit cell have been fully optimized within the PBE+U approach for magnetic ground state (A-AFM).  At each employed $\theta$ angle the atomic positions and the lattice parameters have been fixed to optimal position of magnetic ground state. The direct interband momentum matrix elements were computed from the wave function derivatives using Density Functional Perturbation Theory (DFPT)\cite{PhysRevB.73.045112} in order to determine  the transition dipole strength, as discussed in  \citenum{PhysRevB.101.235408}.

\bibliography{literatura}